\documentclass[aps,amsmath,amssymb,amsfonts,preprint,nofootinbib]{revtex4}

\usepackage[colorlinks=true]{hyperref} 
\hypersetup{
    bookmarks=true,         
    unicode=false,          
    pdftoolbar=true,        
    pdfmenubar=true,        
    pdffitwindow=false,     
    pdfstartview={FitH},    
    pdfsubject={},   
    pdfcreator={},   
    pdfproducer={}, 
    pdfkeywords={} {} {}, 
    pdfnewwindow=true,      
    colorlinks=true,       
    linkcolor=magenta, 
    citecolor=blue,        
    filecolor=magenta,      
    urlcolor=blue           
}

\usepackage{graphicx}

\usepackage{tikz}
\usepackage{caption}
\usepackage{subcaption}
\usepackage{mathtools}

\usetikzlibrary{shapes.misc}
\usetikzlibrary{arrows.meta}
\tikzset{cross/.style={cross out, draw=black, minimum size=2*(#1-\pgflinewidth), inner sep=0pt, outer sep=0pt},
cross/.default={1pt}}
\tikzstyle{axisarrow} = [-{Latex[inset=0pt,length=5pt]}]

\usepackage{calc}

\usepackage{accents}

\newcommand{\ZZ}{{\mathbb Z}}
\newcommand{\RR}{{\mathbb R}}
\newcommand{\ra}{\rightarrow}
\newcommand{\eps}{\epsilon}

\newcommand{\Tr}{{\rm Tr}}
\newcommand{\tr}{{\rm tr}}

\newcommand{\cf}{{\mathcal F}}

\usepackage{amsmath}

\newcommand{\cR}{{\mathcal R}}

\newcommand{\sM}{{\mathsf M}}
\newcommand{\cS}{{\mathcal S}}
\newcommand{\fM}{{\mathfrak M}}

\newcommand{\tH}{{\tilde H}}

\newcommand{\tT}{{\tilde T}}
\newcommand{\tG}{{\tilde G}}

\newcommand{\cm}{{\mathbf m}}

\newcommand{\cM}{{\mathsf M}}

\begin{document}

\title{Higher-dimensional generalizations of the Thouless charge pump}
\author{Anton Kapustin}
\email{kapustin@theory.caltech.edu}
\author{Lev Spodyneiko}
\email{lionspo@caltech.edu}
\affiliation{California Institute of Technology, Pasadena, CA 91125, United States}

\begin{abstract}
We define and study analogs of the Thouless charge pump for many-body gapped systems in dimension $D$. We show how to attach a topological invariant to a $D$-dimensional family of such systems, provided all of them have an on-site $U(1)$ symmetry. For a large class of families we argue that this topological invariant is an integer.  In the case of gapped systems of free fermions in two dimensions, the invariant can be expressed in terms of the curvature of the Bloch-Berry connection. We also obtain a new formula for the Thouless charge pump in 1d which involves only static linear response and is analogous to the Streda formula for Hall conductivity. 
\end{abstract}

\maketitle

\section{Introduction}

It has been argued by D.~Thouless almost 40 years ago \cite{Thouless} that an adiabatic cycling of a gapped 1d system with a $U(1)$ symmetry at zero temperature  results in a quantized charge transport. This is now known as the Thouless charge pump. A simple example is obtained by taking an IQHE system in the cylinder geometry and adiabatically inserting one unit of magnetic flux. This results in a net transport of $\nu$ units of electric charge, where $\nu$ is the number of filled Landau levels. The original argument was stated for gapped systems of non-interacting electrons (possibly with disorder) but later was generalized by Niu and Thouless \cite{NiuThouless} to gapped 1d systems with interactions. They used it to argue that in an FQHE state with a filling fraction $\nu$ the Hall conductance is $\nu$ times $e^2/\hbar$. 

Two key properties of the Thouless charge pump are its integrality and topological invariance (invariance under deformations of the cycle which do not close the energy gap). When attempting to generalize it to higher dimensions, two difficulties present themselves. First, by analogy with the Quantum Hall Effect, one expects that in the presence of nontrivial topological order any topological invariant will be fractional. One expects integrality to hold only for Short-Range Entangled gapped systems. Second, the naive extension of the Thouless charge pump to $D>1$ dimensions is ill-defined, because the charge transported through a $(D-1)$-dimensional hyperplane under adiabatic cycling is typically infinite. If the system is periodic in the directions parallel to the hyperplane, one can consider the charge transported per unit cell. However, in this paper we do not want to assume periodicity. 

Even if the system is not strictly periodic, one expects uniformity on scales much larger than the correlation length. One can try to exploit this to define an analog of the Thouless charge pump in higher dimensions. Consider the case $D=2$ for concreteness and suppose we want to define quantized charge transport in the $x$ direction. One can fix a real number $b$, take the region $b\leq y \leq b+L$ for some $L$ which is much larger than the correlation length, and repeat it periodically in the $y$ direction with period $L$. This gives a system which is periodic in the $y$ direction but perhaps not gapped. Presumably one can fix this by modifying the system near the lines $y=b+nL$ with $n\in\mathbb Z$. Assuming this is done, one can use the gradient expansion to estimate the net charge transported across $x=0$ per unit cell per one cycle as a function of $L$. One expects the leading terms in this expansion to be
\begin{equation}
Q(L)=AL+q+O(1/L).
\end{equation}
For a macroscopically uniform system the coefficient $A$ should be independent of $b$ and of the way we modified the system to make it gapped, but it has dimensions of inverse distance and is not quantized. The dimensionless subleading term $q$ is sensitive to edge effects, such as the choice of $b$ and the details of the modification near $y=b+nL$ and thus cannot be regarded as a characteristic of the original infinite system. It also has no reason to be quantized. 

A possible way out is to consider families of gapped systems depending on $D$ parameters instead of just one. In the above 2d example, suppose that we have a family of gapped systems which depends on two parameters $\lambda_1$ and $\lambda_2$ which both have period $1$. Then we can make $\lambda_1$ to be a function of time $t$ such that $\lambda_1(0)=0$ and $\lambda_1(T)=1$ and make $\lambda_2$ to be a function of $y$ such that $\lambda_2(b)=0$ and $\lambda_2(b+L)=1$. The charge transported per unit cell will have an expansion of the same form as when $\lambda_2$ is fixed:
\begin{equation}
Q'(L)=AL+q'+O(1/L). 
\end{equation}
The coefficient $A$ is expected to be the same, since the effective long-wavelength Hamiltonian differs from the one where $\lambda_2$ is fixed only by terms which are of order $1/L$. The coefficient $q'$ again depends on the choice of $b$ and other details. But we claim that if all the systems in the family have a trivial topological order, the difference $q'-q$ is quantized and is a topological invariant of the family. To see this, imagine stacking the family where $\lambda_2$ varies with $y$  with the time-reversal of the family where $\lambda_2$ is fixed. On the one hand, the net charge transported per unit cell per cycle will be $q'-q+O(1/L)$. On the other hand, considering an $L$-periodic Short-Range Entangled system with a large $L$ is essentially the same as compactifying it on a circle of radius $L$. By the usual quantization of the Thouless charge pump, the charge transported per cycle must be an integer topological invariant. Taking the limit $L\ra\infty$, we conclude that $q'-q$ is an integer topological invariant. Similarly, one can hope to assign an integer topological invariant to a family of Short-Range Entangled $D$-dimensional systems with a $U(1)$ symmetry parameterized by a $D$-dimensional torus $T^D$. 

The primary goal of this paper is to define analogs of the Thouless charge pump for $D$-parameter families of gapped systems with $U(1)$ symmetry in $D$ dimensions. We take a more abstract approach and define topological invariants for families parameterized by arbitrary closed $D$-manifolds, not necessarily having the topology of $T^D$. While the physical interpretation of these topological invariants is not entirely clear, they are very natural from a mathematical standpoint. Let $\fM^{U(1)}_D$ be the space of all gapped $D$-dimensional systems with a $U(1)$ symmetry and a non-degenerate ground state. This is an infinite-dimensional space whose topology is of great interest. For example, its set of connected components is the set of gapped phases with $U(1)$ symmetry in $D$ dimensions. One can view the Thouless charge pump as a map which assigns an integer to a homotopy class of loops in $\fM^{U(1)}_1$. It is clear that this integer is additive under concatenation of loops, so one can view it as a homomorphism from the fundamental group of  $\fM^{U(1)}_1$ to $\ZZ$, or equivalently as an element of $H^1(\fM^{U(1)}_1,\ZZ)$. The existence of cycles with a nonzero Thouless charge pump invariant shows that the space $\fM^{U(1)}_1$ is not simply-connected. To probe higher homotopy or homology groups of spaces $\fM^{U(1)}_D$, it is natural to consider multi-parameter families of gapped systems.\footnote{To avoid possible misconception, we stress that we do not claim that the spaces $\fM^{U(1)}_D$ are simply-connected for $D>1$. We only claim that natural generalizations of the Thouless charge pump to $D$ dimensions cannot detect topologically non-trivial loops in $\fM^{U(1)}_D$ but can detect topologically nontrivial $D$-cycles.} In the context of systems of free fermions, higher-dimensional analogs of the Thouless charge pump have been considered by Teo and Kane \cite{TeoKane}. The existence of higher-dimensional analogs of the Thouless charge pump for families of interacting systems was pointed out by A. Kitaev \cite{Kitaev_unpublished}. 

For technical reasons, we restrict ourselves to the case of lattice systems with a finite-dimensional Hilbert space at each site. This allows us to use the techniques introduced by Kitaev \cite{honeycomb,Kitaev_talk} and developed in our previous papers \cite{ThermalHall, HigherBerry}. 

The content of the paper is as follows. In section II we sketch a field-theoretic argument for the existence of higher-dimensional analogs of the Thouless charge pump. This argument indicates that on $\fM^{U(1)}_D$ there is a closed $D$-form whose cohomology class is independent of any choices. A topological invariant of a $D$-dimensional family of gapped systems can be obtained by integrating this $D$-form over the parameter space of the family. In Section III we review the Thouless charge pump for 1d lattice systems and write down a Kubo-type formula for it. In Section IV we derive an alternative formula for the Thouless charge pump in terms of static linear response. It is analogous to the Streda formula in the theory of the quantum Hall effect and enables to give a one-line proof of the topological nature of the Thouless charge pump. This alternative formula also demonstrates that the Thouless charge pump is a "descendant" of the electric charge in the sense of Ref. \cite{HigherBerry}. In Section V we use the descent formalism to construct higher descendants of the electric charge and use them to define a closed $D$-form on the space $\fM^{U(1)}_D$. Integrals of this closed $D$-form over $D$-cycles are higher-dimensional analogs of the Thouless charge pump. The physical meaning of these invariants for two-dimensional systems is discussed in Section VI. In Section VII we compute the higher Thouless charge pump for families of 2d systems of free fermions and relate it to the curvature of the Bloch-Berry connection. This enables us to give an example of family of gapped 2d systems where our topological invariant is non-trivial. We discuss our results in Section VIII. In an appendix we argue that for a family of $D$-dimensional systems in a trivial phase parameterized by a $D$-sphere our topological invariant is quantized. 

We would like to thank Nikita Sopenko for valuable discussions. This research was supported in part by the U.S.\ Department of Energy, Office of Science, Office of High Energy Physics, under Award Number DE-SC0011632. A.K. was also supported by the Simons Investigator Award.

\section{Effective field theory}

Consider a family of gapped 1d systems parameterized by a manifold $\sM$. If one allows the parameters to vary slowly with space-time coordinates, then the parameters become classical background fields on the two-dimensional space-time $X$. These fields are described by a map $\phi: X\ra\sM$. Integrating out the gapped degrees of freedom, one gets an effective action for these fields. This action may contain topological terms, i.e. terms which do not depend on the metric on $X$. In the context of families of gapped lattice systems such terms have been recently discussed in \cite{HigherBerry}. If all the gapped systems in the family an on-site $U(1)$ symmetry, they can also be coupled to a classical $U(1)$ background gauge field $A$. Then the effective action depends both on $\phi$ and $A$. When $X$ is two-dimensional, gauge-invariance constrains these terms to have the form
\begin{equation}
S_{top}(X,\phi,A)=\int_X \eps^{\mu\nu} A_\mu \partial_\nu\phi^i \tau_i(\phi)d^2 x+\ldots=\int_X A\wedge \phi^*\tau+\ldots,
\end{equation}
where $\tau^i(\phi)$ are components of a closed 1-form $\tau$ on $\sM$, $\eps^{\mu\nu}$ denotes the anti-symmetric tensor density with $\eps^{01}=1$, and dots denote terms independent of $A$. In addition, here and below Greek indices label coordinates on the space-time $X$ while Roman indices label coordinates on the parameter space $\sM$. Varying the above action with respect to $A$, we find a topological term in the current
\begin{equation}\label{topcurrent1d}
J^\mu_{top}(x^0,x^1)=\eps^{\mu\nu}\tau_i\left(\phi(x^0,x^1)\right)\partial_\nu\phi^i(x^0,x^1).
\end{equation}
Such topological terms in the $U(1)$ current have been discovered by Goldstone and Wilczek in their work on soliton charges \cite{GoldstoneWilczek}.
If we now let $\phi^i$ be independent of $x^1$ and be periodic functions of $x^0$ with period $T$, we find that the net topological charge per period transported though a section $x^1=a$ is given by 
\begin{equation}
\Delta Q (a)=\int_0^T J^1_{top}(x^0,a) dx^0=-\oint \tau_i(\phi) d\phi^i . 
\end{equation}
We see that the topological charge transport is independent of $a$ and given by an integral of $\tau$ along the corresponding loop in $\sM$. Such an integral is called a period of the 1-form $\tau$. A period of a closed 1-form does not change under continuous deformations of the loop, so $\Delta Q(a)$ is a topological invariant.

To see that $\Delta Q $ is quantized, one needs to exploit the invariance of the effective field theory under "large" $U(1)$ gauge transformation which cannot be deformed to a constant. Such a transformation can be described by a continuous multi-valued function $f:X\ra \RR$ whose values are defined up to an integer multiple of $2\pi$. Performing the gauge transformation $A\mapsto A+df$, we find that the action changes by $\int_X df\wedge \phi^*\tau$. For $e^{iS_{top}}$ to be unchanged, the change in the action must be an integral multiple of $2\pi$. This is satisfied for all conceivable $f$ and $\phi$ if and only if integrals of $\tau$ over arbitrary 1-cycles on $\sM$ (that is, all periods of $\tau$) are integers. Therefore $\Delta Q $ is also an integer.

In higher dimensions, gauge-invariance allows for a more complicated dependence of $S_{top}$ on $A$. For example, quantum Hall response for 2d systems is described by the Chern-Simons action. However, if our goal is to generalize the Thouless charge pump to higher dimensions, we can focus on the terms which  modify the current even when the gauge field strength vanishes. Such terms are linear in $A$ and have the form
\begin{equation}
S_{top}(X,A,\phi)=\int_X A\wedge\phi^*\tau,
\end{equation}
where $X$ is $(D+1)$-dimensional space-time, $\phi:X\ra \sM$ is a smooth map, and  $\tau$ is a closed $D$-form on $\sM$. Invariance with respect to "large" gauge transformations again imposes restrictions on periods of $\tau$ (that is, integrals of $\tau$ over $D$-cycles in $\sM$). It is not entirely obvious what these restrictions are for general $D$. To see what the difficulty is, consider the current corresponding to the above action:
\begin{equation}
J^\mu_{top}=\eps^{\mu\nu_1\ldots\nu_D}\tau_{i_1\ldots i_D} \partial_{\nu_1}\phi^{i_1}\ldots \partial_{\nu_D}\phi^{i_D}.
\end{equation}
One effect of such a topological term in the current is to give charge to "skyrmions", i.e. topologically non-trivial configurations of fields $\phi^i$. Suppose the space-time has the form $Y\times \RR$, where the spatial manifold $Y$ is closed and oriented. Then the charge of a skyrmion is $Q_Y(\phi)=\int_Y J=\int_Y\phi^*\tau$. For $Q_Y(\phi)$ to be integral, $\tau$ must integrate to an integer over any $D$-cycle on $\sM$ of the form $\phi_*[Y]$, where $[Y]$ is the fundamental class of the $D$-manifold $Y$, and $\phi_*$ denotes the pushforward map in homology. But for a  general manifold $\sM$ and a general $D$ it is not true that any $D$-cycle in a manifold $\sM$ can be realized as a pushforward of the fundamental homology class of some closed  oriented $D$-manifold \cite{Thom}. In the fermionic case the geometry of $Y$ is even more restricted (it must be a $Spin^c$ manifold \cite{Kapustinetal}), so for general $D$ the integrality of charge is not equivalent to the integrality of periods of $\tau$. Nevertheless, for sufficiently low $D$ the integrality of charge does require the periods of $\tau$ to be integral. In the bosonic case, this follows from the fact that for $D\leq 6$ any $D$-cycle is a pushforward of the fundamental homology class of a closed oriented $D$-manifold \cite{Thom}. Since any closed oriented $D$-manifold with $D\leq 3$ has a spin structure, for fermionic systems in dimension $D\leq 3$ the integrality of charge also requires the integrality of periods of $\tau$.  Also, if the parameter space is a $D$-dimensional sphere or a $D$-dimensional torus, one can set $Y=S^D$ or $Y=T^D$ both in the bosonic and fermionic cases. In this case $\tau$ is a top-degree form, and the skyrmion charge is its integral over the whole $\sM$. Thus the integrality of charge requires the integral of $\tau$ over $S^D$ or $T^D$ to be integral.

The argument for quantization of $\tau$ in higher dimensions has an important loophole. If the gapped system of interest is in a topologically ordered phase, the effective field theory should really be a TQFT coupled to $\phi$ and $A$. In this situation $S_{top}$ written above need not be invariant under large gauge transformations by itself, and accordingly its integrals over $D$-cycles need not be integral. 

\section{Thouless charge pump for 1d lattice systems}\label{section: 1d thouless}

Consider a lattice system in one spatial dimension, with a finite-dimensional on-site Hilbert space and an on-site $U(1)$ symmetry. In more detail, the lattice $\Lambda$ is an infinite subset of $\RR$ which is discrete and without accumulation points, the Hamiltonian has the form $H=\sum_{p\in\Lambda} H_p$, where each $H_p$ has a range no larger than some fixed $R>0$, and the charge operator $Q_{tot}$ has the form $Q_{tot}=\sum_{p\in\Lambda} Q_p$, where each $Q_p$ has integer eigenvalues and acts only on degrees of freedom at site $p$. Invariance of $H$ under $U(1)$ symmetry means that $[Q_{tot},H_p]=0$ for all $p$. We also assume that there is a unique ground state, with a nonzero energy gap to excited states. If each $H$ depends on a parameter $\lambda$, we will say that we have a one-parameter family of gapped $U(1)$-invariant 1d systems. We will assume that $\lambda$ is periodically identified with period $1$, so that our one-parameter family is a loop $\Gamma$ in the space of gapped $U(1)$-invariant 1d systems. 

Following Thouless \cite{Thouless}, we consider slowly varying $\lambda$ as a function of time $t$, so that $\lambda(0)=0$ and $\lambda(T)=1$ for some large time $T$. This will drive the system out of the ground state and create nonzero charge current. This current can be found as follows \cite{NiuThouless}. 

We can separate the density matrix into an instantaneous part and a small deviation:
\begin{equation}
    \rho(t ) = \rho_{inst} (\lambda(t)) + \Delta \rho (t),
\end{equation}
where 
\begin{equation}
    \rho_{inst} (\lambda)= |0(\lambda) \rangle \langle 0(\lambda)|
\end{equation}
and $|0(\lambda)\rangle$ is the ground state of Hamiltonian $H(\lambda)$.   We may always normalize the lowest eigenvalue to be $0$, for all $\lambda$.

The equation of motion is 
\begin{equation}
    -i[H,\Delta\rho] =  -i[H,\rho_{inst}+\Delta\rho]  = \dot\rho_{inst} + \Delta \dot \rho \approx \dot\rho_{inst},
\end{equation}
where dot is derivative with respect to $t$ and  we have dropped  $\Delta \dot \rho$ since it is small in the adiabatic approximation. Sandwiching this equation between instantaneous ground state $ |0(\lambda)\rangle$ with energy 0 and instantaneous excited state $|n(\lambda)\rangle$ with energy $E_n(\lambda)$, we find
\begin{align}\label{delta rho}
    \langle 0 | \Delta \rho |n\rangle  = -\frac{i\langle \dot 0 | n\rangle } {E_n} = \frac{i\langle  0 |\dot H| n\rangle } {E_n^2}.
\end{align}
Here in the last step we used the perturbation theory formula
\begin{equation}
 |\dot 0\rangle = \dot \lambda \dfrac{\partial}{\partial \lambda} | 0\rangle = -\dot \lambda \frac{1}{H}P \dfrac{\partial H}{\partial \lambda} | 0\rangle =  -  \frac{1}{H} P \dot H| 0\rangle ,
\end{equation}
where $P$ is the projector to the excited states.

Let $J(a)$ be the operator corresponding to the current through a point $a\in\RR$ (its precise form will be discussed shortly). The total charge passing through the point $a$ over one period $T$ is given by
\begin{equation}
    \Delta Q = \int_0^T dt\left[ (1+\Delta\rho_{00}) J(a)_{00} +\sum_{n\ne0} (\Delta \rho_{0n} J(a)_{n0}+J(a)_{0n}\Delta \rho_{n0} ) \right],
\end{equation}
where we have dropped terms which are small in the adiabatic expansion. 
Due to Bloch's theorem \cite{Bohm,WatanabeBloch}, the net current in a ground state is zero, hence $J(a)_{00}=0$. The remaining terms can be rewritten using (\ref{delta rho}) as 
\begin{align}
    \Delta Q =  i\int_0^T dt\left[ \langle \dot H \frac{P}{H^2} J(a) \rangle - \langle J(a)\frac{P}{H^2}  \dot H \rangle  \right] =i\int_0^T dt\oint  \frac{dz}{2\pi i} \Tr \left(G \dot H G^2 J(a) \right)  .
\end{align}
Here $G=(z-H)^{-1}$ is the many-body Green's function, and the contour integral in the $z$-plane is taken along a loop surrounding only the lowest eigenvalue of $H$.

The integral over time $t$ can be replaced with an integral in the parameter space. The result is that the total charge transported during periodic adiabatic variation is given by 
\begin{equation}\label{Kubo}
\Delta Q=  i \int_0^1 d\lambda \oint  \frac{dz}{2\pi i} \Tr \left(G \frac{dH}{d\lambda} G^2 J(a) \right).
\end{equation}

In the case of lattice systems, the current from site $q$ to site $p$ is given by 
\begin{equation}
J_{pq}=i[H_q,Q_p]-i[H_p,Q_q],
\end{equation}
which measures the current from $q$ to $p$. The current operator is constructed so that the conservation equation
\begin{equation}\label{electric conservation law}
\frac{dQ_q}{dt}=i\sum_{p\in\Lambda} [H_p,Q_q] = - \sum_{p\in\Lambda}J_{pq},
\end{equation}
is automatically satisfied. Under some natural assumptions this choice of the current operator is essentially unique (see \cite{ThermalHall} for more details).
Then the current through a point $a\notin\Lambda$ is given by
\begin{equation}
J(a)=\sum_{ p>a  \atop q<a} J_{pq}.
\end{equation}

\section{A static formula for the Thouless charge pump}

As explained in \cite{Thouless}, Hall conductance can be viewed as a special case of the charge pump. For Hall conductance, there are two types of formulas: the Streda formula \cite{Streda} and various versions of the Kubo formula (see e.g. \cite{NiuThouless}). The Streda formula at zero temperature involves only static linear response. The Kubo formula involves dynamic response even if one specializes to zero temperature and is more subtle. In this section we derive a formula for the Thouless charge pump which involves only static linear response and thus is analogous to the zero-temperature Streda formula . It turns out to be a more convenient starting point for higher-dimensional generalizations.

As a warm-up, consider a family of gapped 0d quantum-mechanical systems with a $U(1)$ symmetry and a non-degenerate ground state parameterized by a manifold $\sM$. We will collectively denote the parameters $\lambda^\ell$ as $\lambda$. The charge operator $Q$ has integer eigenvalues and is assumed to be independent of the parameters. This is because the symmetry action on the Hilbert space is fixed. The Hamiltonian $H$ is a Hermitian operator continuously depending on the parameters $\lambda$. By adding to $H(\lambda)$ a scalar depending on $\lambda$, one can normalize the ground-state energy to be zero for all $\lambda$. The ground-state charge $Q^{(0)}=\langle Q\rangle$ is independent of $\lambda$ because it is an integer and varies continuously with $\lambda$. One can also prove this without using integrality: 
\begin{align}
    d\langle Q\rangle = -\left\langle dH \frac P H  Q\right\rangle-\left\langle Q \frac P H  dH \right\rangle =-Q^{(0)}\left\langle dH \frac P H  \right \rangle-Q^{(0)}\left\langle  \frac P H  dH \right\rangle =0. 
\end{align}
Here $P$ is the projector to excited states, $d = \sum_{\ell}^{} d\lambda^\ell \frac{\partial }{\partial \lambda^\ell}$ is the exterior differential on $\sM$, and the angular brackets denote ground-state average. 

Turning to gapped many-body systems in dimension $D>0$, we note that the Goldstone theorem implies that the $U(1)$ symmetry is unbroken, and thus expectation values of the form $\langle [Q_{tot},A]\rangle$ vanish for all local operators $A$. Here $Q_{tot}=\sum_p Q_p$ as before. This does not mean, however, that the ground-state is annihilated by $Q_{tot}$. The operator $Q_{tot}$ is unbounded, and its ground-state expectation value is typically ill-defined. The change in the expectation value of $Q_{tot}$ under variation of $\lambda$ is typically ill-defined too.

For $D=1$ one can hope that the change in the expectation of the charge on the half-line $p>a$ is finite. Indeed, one expects this to be equal to the charge which flows from the region $p<a$ to the region $p>a$ as one changes parameters. Since the current operator $J(a)$ is bounded for $D=1$, the change in the charge should be well-defined. Some regularization might be needed though. 

The infinitesimal change in the expectation value of the charge $Q_q$ at site $q$ can be computed using static linear response theory:
\begin{equation}\label{changeQq}
d\langle Q_q\rangle= \oint \frac{dz}{2\pi i} \Tr \left( G dH G Q_q\right).
\end{equation}
Here we used an integral representation of the projector to the ground state $1-P$:
\begin{equation}
1-P=\oint \frac{dz}{2\pi i} \frac{1}{z-H}.
\end{equation}
The expression (\ref{changeQq}) is well-defined because one can write it as an absolutely convergent sum of correlators of local observables:
\begin{equation}
d\langle Q_q\rangle=\sum_{p\in\Lambda} \oint \frac{dz}{2\pi i} \Tr \left( G dH_p G Q_q\right)
\end{equation}
Indeed, the terms in this sum decay exponentially with $|p-q|$ \cite{Watanabe}. On the other hand, the sum $\sum_{q>a} d\langle Q_q\rangle $ has no reason to be absolutely convergent and its value is ambiguous. To make sense of it, we first of all rewrite eq. (\ref{changeQq}) as follows:
\begin{equation} \label{1d descent eq}
d \langle  Q_q \rangle =\sum_{p\in \Lambda}T^{(1)}_{pq},
\end{equation}
where
\begin{equation}\label{T1}
 T^{(1)}_{pq} = \oint \frac{dz}{2\pi i} \Tr \left( G dH_pG Q_q -G dH_q G Q_p  \right).
\end{equation}
The second term in $T^{(1)}_{pq}$ gives zero contribution to $d \langle  Q_q \rangle $, since 
\begin{equation}
-\oint \frac{dz}{2\pi i} \Tr \left( G dH_q G Q_{tot} \right)=-\oint \frac{dz}{2\pi i} \Tr \left( G^2 dH_q Q_{tot}  \right)=\oint \frac{dz}{2\pi i} \frac{\partial}{\partial z} \Tr \left( G dH_q Q_{tot}  \right)=0.
\end{equation}
Introducing $f(p)=\theta(p-a)$ and using the skew-symmetry of $T^{(1)}_{pq}$, one can formally write
\begin{equation}
\sum_{q>a} d\langle Q_q\rangle =\sum_{p,q\in\Lambda} f(q) T^{(1)}_{pq}=\frac12 \sum_{p,q\in\Lambda} (f(q)-f(p)) T^{(1)}_{pq}.
\end{equation}
The rightmost sum is absolutely convergent because $T^{(1)}_{pq}$ decays exponentially when $|p-q|$ is large \cite{Watanabe}, while $f(q)-f(p)$ is nonzero only when $q>a$ and $p<a$ or $q<a$ and $p>a$. To make the convergence more obvious one can write the rightmost sum more explicitly as $\sum_{ p<a  \atop q>a}T^{(1)}_{pq}$.

Let us therefore define a 1-form $Q^{(1)}(f)$ on the parameter space of gapped 1d systems with a $U(1)$ symmetry
\begin{equation}\label{Q1d}
Q^{(1)}(f)=\frac12 \sum_{p,q\in\Lambda} (f(q)-f(p)) T^{(1)}_{pq}.
\end{equation}
It will be convenient to allow $f$ to be an arbitrary real function on the lattice $\Lambda$ such that $f(p)=0$ for $p\ll 0$ and $f(p)=1$ for $p\gg 0$. Convergence still holds, as can be easily verified. In the case $f(p)=\theta(p-a)$ the 1-form $Q^{(1)}(f)$ has the meaning of the regularized differential of the charge on the half-line $p>a$. 

We are now going to relate $Q^{(1)}(f)$ to the Thouless charge pump. Eq. (\ref{Kubo}) expresses the net charge pumped during one cycle as an integral of a 1-form on the parameter space. We would like to compare this 1-form with the 1-form $Q^{(1)}(f)$. The first step is to generalize the 1-form appearing in (\ref{Kubo}) by allowing dependence on a function $f:\Lambda\ra\RR$. This is achieved by replacing the current operator $J(a)$ with a smeared current operator 
\begin{equation}
-J(\delta f)=-\frac12 \sum_{p,q\in\Lambda} (f(q)-f(p)) J_{pq}.
\end{equation}
If we assume $f(p)=1$ for $p\gg 0$ and $f(p)=0$ for $p\ll 0$, then this expression is well-defined. If we set $f(p)=\theta(p-a)$, it reduces to $J(a)$. Ignoring convergence issues, one can formally write $J(\delta f)=\sum_{p,q} f(q) J_{pq}$. Upon using the conservation law (\ref{electric conservation law}), this becomes $-d Q(f)/dt$, where $Q(f)=\sum_q f(q) Q_q$ is the smeared charge on a half-line. Of course, these manipulations are formal, since for $f$ as above both the operator $Q(f)$ and its time-derivative is unbounded, and so is its time-derivative. In any case, replacing $J(a)$ with $-J(\delta f)$ we get a 1-form on the parameter space
\begin{equation}
{\tilde Q}^{(1)}(f)=-i \oint \frac{dz}{2\pi i} \Tr \left( G dH_q G^2 J(\delta f) \right).
\end{equation}
We provisionally denoted this 1-form $\tilde Q^{(1)}(f)$, but in fact it coincides with $Q^{(1)}(f)$. Indeed, their difference is
\begin{align}
    \begin{split}
     &\frac 1 2 \sum_{p,q\in\Lambda} (f(q)-f(p))  \oint \frac{dz}{2\pi i} \Tr \left( G dH_pG Q_q -G dH_q G Q_p +iG dH G^2 J_{pq} \right)\\&=\frac 1 2 \sum_{p,q,r\in\Lambda} (f(q)-f(p))  \oint \frac{dz}{2\pi i} \Tr \left( i G dH_pG^2 J_{qr}+iG dH_q G^2 J_{rp} +iG dH_r G^2 J_{pq} \right)\\&=\frac 1 6 \sum_{p,q,r\in\Lambda} (f(q)-f(p)+f(r)-f(q)+f(p)-f(r)) \\ &\qquad \qquad \qquad \qquad\times\oint \frac{dz}{2\pi i} \Tr \left( i G dH_pG^2 J_{qr}+iG dH_q G^2 J_{rp} +iG dH_r G^2 J_{pq} \right)=0.
    \end{split}
\end{align}
Thus we proved a new formula for the Thouless charge pump:
\begin{equation}
\Delta Q=\int Q^{(1)}(f).
\end{equation}
Here  $Q^{(1)}(f)$ is a 1-form on the parameter space $\sM$ and the integration is over a loop in $\sM$ specifying the periodic family of Hamiltonians we are interested in. In principle one should set $f(p)=\theta(p-a).$ In fact, this expression is independent of $f$, provided the asymptotic behavior is as required above. Indeed, given any other such function $f'$, the difference $g=f'-f$ is compactly supported. Therefore
\begin{equation}\label{q1 compactly supported}
Q^{(1)}(f')-Q^{(1)}(f)=\frac12 \sum_{p,q}(g(q)-g(p)) T^{(1)}_{pq}=\sum_{p,q} g(q) T^{(1)}_{pq}=d\langle Q(g)\rangle . 
\end{equation}
Here $Q(g)=\sum_q g(q) Q_q$. Since the difference of these two 1-forms is exact, their integrals over any loop are the same. 

We note the following useful properties of the 1-forms $T^{(1)}_{pq}$ and $Q^{(1)}(f)$. Suppose we replace each $H_p$ with its complex-conjugate $H_p^*$. Physically this corresponds to time-reversal. Using the Hermiticity of $H_p$ and $Q_p$, it is easy to see that this operation maps $T^{(1)}_{pq}$ to $T^{(1)}_{qp}$ and thus reverses the sign of $Q^{(1)}(f)$. This agrees with the intuition that time-reversal flips the sign of the Thouless charge pump. Another useful fact is when we stack together two independent families of systems, the corresponding 1-forms $T^{(1)}_{pq}$ and $Q^{(1)}(f)$ add up. This is not easy to see from the above formulas. However, it follows from the physical interpretation of the Thouless charge pump and can be shown formally by re-writing the above formulas in terms of Kubo's canonical correlation function.

The new formula is also a convenient starting point to for proving that the Thouless charge pump is a topological invariant, i.e. that it does not change under continuous deformations of the loop in the parameter space. The integral of a 1-form does not vary under continuous changes of the contour if and only if the 1-form is closed. As we will show later in the paper, there exists a 2-form $T^{(2)}_{pqr}$ which is skew-symmetric in $p,q,r$, decays exponentially when any of the pairwise distances between $p,q,r$ are large, and satisfies 
\begin{align}
    dT^{(1)}_{qr}= \sum_{p\in\Lambda} T^{(2)}_{pqr}.
\end{align}
This can be used to show that $dQ^{(1)}(f)=0$:
\begin{align}
    \begin{split}
    dQ^{(1)}(f)= \frac 1 2 \sum_{p,q\in \Lambda}(f(q)-f(p)) dT^{(1)}_{pq}= \frac 1 2 \sum_{p,q,r\in \Lambda}(f(q)-f(p)) T^{(2)}_{pqr} \\= \frac 1 6 \sum_{p,q,r\in \Lambda}(f(q)-f(p)+f(r)-f(q)+f(p)-f(r)) T^{(2)}_{pqr}=0.        
    \end{split}
\end{align}

Putting the above observations together, we conclude that $Q^{(1)}(f)$ defines a degree-1 cohomology class on the parameter space which does not depend on the choice of $f$. The Thouless charge pump for any loop is the evaluation of this cohomology class on the homology class of the loop. This makes the topological nature of the Thouless charge pump completely explicit.
In Appendix \ref{appendix:quantization} we argue that the integral of $Q^{(1)}(f)$  over any loop in the parameter space is an integer.
So in fact the cohomology class of $Q^{(1)}(f)$ is integral and $\Delta Q$ is quantized.

\section{Descendants of the Thouless charge pump}

Trying to keep the same level of transparency in formulas is hard when passing to higher dimensions. The formulas tend to be clumsy and computations become cumbersome. However, introduction of some straightforward and natural notations makes the equations look elegant and cumbersome computations become a simple routine. In this section, we will use notations and definitions explained in Appendix B of \cite{ThermalHall} (see also Section IV of \cite{HigherBerry} where this machinery is introduced for a similar problem).

A lattice in dimension $D$ is an infinite  subset $\Lambda\subset\RR^D$ which is discrete and has no accumulation points. Later we will impose further constraints on $\Lambda$ ensuring that $\Lambda$ fills the whole $\RR^D$ uniformly. The Hamiltonian has the form $H=\sum_{p\in\Lambda} H_p$, where each $H_p$ is a local observable localized on some neighborhood of $p\in\Lambda$. We will assume that the size of this neighborhood is bounded from above by some $R>0$. That is, we will only consider finite-range Hamiltonians. The charge operator $Q_{tot}$ has the form $Q_{tot}=\sum_{p\in\Lambda} Q_p$, where $Q_p$ is localized at $p$ and has integral eigenvalues. A Hamiltonian is said to be $U(1)$-invariant if $[H_p,Q_{tot}]=0$ for all $p\in\Lambda$.  A $n$-chain on $\Lambda$ is a function of $n+1$ points of $\Lambda$ which is skew-symmetric under their permutations and satisfies certain decay conditions when any two points are far from each other, see Refs.  \cite{ThermalHall,HigherBerry} for details. For example, $H_p$ and $Q_p$ are operator-valued 0-chains, while the current $J_{pq}$ is an operator-valued 1-chain. There is a natural operation on chains $\partial$ which decreases the degree of a chain by $1$ and satisfies $\partial^2=0$. For example, if $J$ is a 1-chain, the corresponding 0-chain $\partial J$ has components $(\partial J)_q=\sum_p J_{pq}$. The reader is advised to consult Refs.  \cite{ThermalHall,HigherBerry} for details. 

We have seen that the Thouless charge pump of a 1d system can be expressed as an integral of a 1-form on the parameter space. This 1-form represents a suitably regularized differential of the ground-state charge on a half-line. The ground-state charge of the whole 1d system as well as its differential are ill-defined in the infinite volume limit. In a sense, the Thouless charge pump is a 1d descendant of the ground-state charge. Mathematically, this statement is encapsulated by eq. (\ref{1d descent eq}). Using the chain notation, it can be written as
\begin{equation}
    d T^{(0)}  = \partial T^{(1)},
\end{equation}
where $T^{(0)}$ is a 0-chain with components $\langle Q_p\rangle$, $T^{(1)}$ is 1-chain with components $T^{(1)}_{pq},$ $d$ is the exterior  differential on the parameter space, and $\partial$ is the boundary operator on chains. Note that the chain $T^{(0)}$ takes values in 0-forms on the parameter space while the chain $T^{(1)}$ takes values in 1-forms on the parameter space. If we say that the total degree of an $n$-chain with values in $m$-forms is $n-m$, then both $T^{(0)}$ and $T^{(1)}$ have total degree $0$. Both $d$ and $\partial$ increase the total degree by $1$.

Generalising this we can consider a more general ``descent equation'':
\begin{equation}
    d T^{(n)} = \partial T^{(n+1)},
\end{equation}
with $T^{(n)}$ is an $n$-chain with value in $n$-forms on the parameter space. This descent equations was first proposed by A. Kitaev \cite{Kitaev_talk,Kitaev_unpublished}.
Using the 0-chain $T^{(0)}$ with components $\langle Q_p\rangle$ as initial condition, a solution to this equation can be found to be
\begin{equation}\label{tn expr}
    T^{(n)}_{p_0\dots p_n} =\sum_{\sigma \in {\mathcal S}_{n+1}} (-1)^{{\rm sgn}\, \sigma}\oint \frac{dz}{2\pi i} \Tr \left( G dH_{p_{\sigma(0)}}G dH_{p_{\sigma(1)}}\dots G dH_{p_{\sigma(n-1)}} G Q_{p_{\sigma(n)}} \right),
\end{equation}
where ${\mathcal S}_{n+1}$ is the permutation group on $n+1$ objects and wedge product of forms $\wedge$ is implicit. For $n=2$ the expression $T^{(2)}_{pqr}$ decays exponentially when $|p-q|$, $|q-r|$, or $|p-r|$ are large \cite{Watanabe}. The physical interpretation is that static linear response of a gapped system to perturbations near a point $p$ is insensitive to variations of the Hamiltonian far from $p$. We expect that the method of \cite{Watanabe} can be used to show that $T^{(n)}$ for any $n$ decays exponentially when any two of the points $p_0,\ldots,p_n$ are far apart. This means that $T^{(n)}$ is a well-defined $n$-chain. Physically, such a decay means that the all-order non-linear static response of a gapped  system to perturbations near a point $p$ is insensitive to variations of the Hamiltonian far from $p$.

Note that the descent equation for $n=2$ was used by us to argue that the 1-form $Q^{(1)}(f)$ is closed. Now that we know that a solution to the $n=2$ descent equation exists, our proof that $Q^{(1)}(f)$ is closed is complete.

In order to get an $n$-form on the parameter space we need to contract the $n$-chain $T^{(n)}$ with some $n$-cochain $\alpha$. An $n$-cochain is a function of $n+1$ points of $\Lambda$ which is skew-symmetric under permutations of the points and satisfies certain decay conditions spelled out in \cite{ThermalHall,HigherBerry}. Roughly, these decay conditions say that when all points go off to infinity while staying within some finite distance of each other, the function eventually becomes zero. A natural operation on cochains $\delta$ increases the degree of a cochain by one and satisfies $\delta^2=0$. An $n$-chain $A(p_0,\ldots,p_n)$ can be evaluated on an $n$-cochain $\alpha(p_0,\ldots,p_n)$ to give a number which we denote $\langle A,\alpha\rangle$. By definition, 
\begin{equation}
\langle A,\alpha\rangle=\frac{1}{(n+1)!}\sum_{p_0,\ldots,p_n} A(p_0,\ldots, p_n)\alpha(p_0,\ldots,p_n). 
\end{equation}
The operations $\partial$ and $\delta$ are dual to each other in the sense that for any $n$-chain $A$ and any $(n-1)$-cochain $\beta$ one has an identity $\langle \partial A,\beta\rangle=\langle A,\delta\beta\rangle.$ 

To get an $n$-form on $\sM$ it is natural to consider expressions of the form $\langle T^{(n)},\alpha\rangle $ for some $n$-cochain $\alpha$. An $n$-form can be integrated over a closed $n$-manifold in the parameter space, or more generally over an $n$-cycle in the parameter space. Such an integral is a topological invariant (unchanged under deformations of the cycle) if and only if the $n$-form $\langle T^{(n)}, \alpha \rangle$ is closed.  Using the descent equation we get:
\begin{align}
    d\langle T^{(n)}, \alpha \rangle = \langle d T^{(n)}, \alpha \rangle = \langle \partial T^{(n+1)}, \alpha \rangle  = \langle  T^{(n+1)}, \delta\alpha \rangle.
\end{align}
Thus we want the cochain $\alpha$ to be closed, $ \delta\alpha=0$. Also, for exact cochain $\alpha = \delta \gamma $ we find
\begin{equation}
    \langle T^{(n)}, \alpha \rangle=\langle T^{(n)}, \delta \gamma  \rangle = \langle \partial T^{(n)}, \gamma  \rangle =  \langle d T^{(n-1)}, \gamma  \rangle=  d\langle  T^{(n-1)}, \gamma  \rangle.
\end{equation}
This means that integral of such a form over any cycle will be zero. 

Combining these two facts we see that in order to get a non-trivial topological invariant one has to contract the $n$-chain $T^{(n)}$ with an $n$-cochain $\alpha$ which is closed but not exact. As explained in \cite{Roe}, such cochains are controlled by the coarse geometry of $\Lambda$. If $\Lambda\subset\RR^D$ is coarsely equivalent to $\RR^D$ (in the sense that there exists some $R>0$ such that any point of $\RR^D$ is within distance $R$ of some point of $\Lambda$), then one can construct closed but not exact cochains of degree $D$, but not in other degrees. Moreover, the space of closed $D$-cochains modulo exact $D$-cochains (that is, degree-$D$ cohomology) is one-dimensional, so up to an overall scaling $\alpha$ is essentially unique. One can take it to be $\delta f_1 \cup \dots \cup \delta f_D$ where $f_{\mu}= \theta(x^\mu(p))$ with $x^\mu(p)$ being the $\mu$-coordinate of cite $p$, and $\cup$ is a product operation on cochains, see Refs. \cite{ThermalHall,HigherBerry}.

The conclusion is that one can define a topological invariant of $D$-parameter families of gapped $D$-dimensional systems as an integral of a $D$-form $Q^{(D)}(f_1,\ldots,f_D)$ on the parameter space. This $D$-form is given by 
\begin{align}
    Q^{(D)}(f_1,\dots,f_D) = \langle T^{(n)}, \delta f_1 \cup \dots \cup\delta f_D\rangle. 
\end{align}
The integral of this $D$-form over a $D$-cycle in the parameter space is invariant under deformations of the cycle. It is also invriant under  changes of the functions $f_\mu(p)$, provided they have the same asymptotic behavior as $\theta(x^\mu(p))$. Thus integrals of $Q^{(D)}$ over $D$-cycles in the parameter space are a natural generalization of the Thouless charge pump to dimension $D$. 

We argue in Appendix \ref{appendix:quantization} that the form $Q^{(D)}$ is properly normalized, in the sense that for familes of Short-Range Entangled systems its integrals over spherical cycles are integer.

\section{Physical interpretation of the 2d Thouless charge pump}

Our approach to defining higher-dimensional analogs of the Thouless charge pump was rather formal. In this section we are going to clarify their physical meaning in the case when the system is two-dimensional and the parameter space is a torus of dimension two. As proposed in the introduction, the physical interpretation involves making the parameters of the Hamiltonian slowly varying functions of both time and spatial coordinates. 

As a warm-up, let us discuss an alternative interpretation of the  usual Thouless charge pump for gapped 1d systems. As discussed in Section 2, the same term in the effective  action gives rise to the Thouless charge pump and gives charge to 1d skyrmions. A skyrmion is defined as a topologically nontrivial map $\phi$ from $\RR$ to the parameter space $\sM$ which approaches the same point  both at $x=-\infty$ and $x=+\infty$. Such a map is topologically the same as a loop in the parameter space with a basepoint corresponding to the value of the parameters at $x=\pm\infty$. It follows from eq. (\ref{topcurrent1d}) that the topological invariant $\Delta Q=\int \phi^*\tau$ attached to a loop can be interpreted in two different ways: as minus the net charge pumped through $x=0$ when the loop parameter is a slowly-varying function of time and as the charge of a skyrmion corresponding to the loop. From the point of view of lattice models, it is far from obvious that the same topological invariant controls both quantities. Our first goal is to show that this is indeed the case.

Let us consider the parameter space $\sM$ given by a loop $S^1$ parameterized by a variable $\lambda\in [0,1]$ such that both $0$ and $1$ correspond to the basepoint. Thus we have a one-parameter family of gapped $U(1)$-invariant Hamiltonians 
\begin{equation}
H(\lambda)=\sum_p H_p(\lambda) 
\end{equation}
such that $H_p(0)=H_p(1)$. We assume that all these Hamiltonians have a unique ground state and denote by $\xi$ the supremum of the correlation lengths of these ground states.

Let $g:\RR\ra [0,1]$ be a continuous function defined as follows:
\begin{equation}\label{function g}
g(x)=\left\{\begin{array}{ll} 0, & x<L \\ \frac{x}{L}-1, & L\leq x\leq 2L \\ 1, & x>2L \end{array}\right.
\end{equation}
The "skyrmion" Hamiltonian $H^s$ is obtained by making $\lambda$ depend on $p$:
\begin{equation}
H^s=\sum_p H^s_p=\sum_p H_p\left(g(p)\right).
\end{equation}
Thus $H^s_p=H_p(0)$ if $p<L$ or  $p>2L$.

To proceed, we need to make a technical assumption. Suppose we are given a family of gapped Hamiltonians  depending on some parameters which live in a compact parameter space $\cR$. Suppose also that all these Hamiltonians have a unique ground state and let $\xi$ be the supremum of the correlation lengths of all the ground states. Now suppose we make the parameters slowly varying functions of coordinates. By a slow variation we mean that the parameters vary appreciably over a scale $L$ which is much larger than $\xi$. Our technical assumption will be that the new Hamiltonian is still gapped, has a unique ground state, and its correlation length is of order $\xi$ and thus is still much smaller than $L$. Essentially, this is the same as assuming that derivative expansion makes sense for gapped systems with a unique ground state. It is very natural from the physical viewpoint and plays an important role in many arguments pertaining to topological invariants of gapped systems (see e.g. \cite{ThermalHall,HigherBerry}). It would be very desirable to find a rigorous proof of this assumption.

Assuming this, we can apply the results of \cite{Watanabe} on the insensitivity of the expectation values of local observables on the behavior of the Hamiltonian far from the support of the observable. In particular, the expectation values of observables supported at $x<0$ or  $x>3L$ in the ground state of $H^s$ are exponentially close to the expectation values of the same observables in the ground state of $H(0)$. Therefore we can define the skyrmion charge as follows:
\begin{equation}
Q^s=\lim_{L\ra\infty}\sum_p\left( \langle Q_p\rangle_s-\langle Q_p\rangle_0\right),
\end{equation}
where $\langle \ldots\rangle_s$ and $\langle\ldots \rangle_0$ denote expectation values in ground states of $H^s$ and $H(0)$, respectively. The sum over $p$ is converging exponentially fast away from $L<p<2L$, so we can write
\begin{equation}\label{tildeQ}
Q^s=\sum_p \langle Q_p\rangle_s h(p)-\sum_p \langle Q_p\rangle_0 h(p)+O(L^{-\infty}),
\end{equation}
where $h(x)=\theta(x)-\theta(x-3L)$.

To compute the r.h.s. of eq. (\ref{tildeQ}) we need a family of gapped Hamiltonians interpolating between $H(0)$ and $H^s$. Since the loop used to define $H^s$ is assumed to be non-contractible, such an interpolation   does not exist if we require the asymptotic behavior at $x=\pm\infty$ to be fixed. But if we relax this constraint, the difficulty disappears. We are going to use the following one-parameter family:
\begin{equation}
\tH(\mu)=\sum_p \tH_p(\mu)=\sum_p H_p(g_\mu(p)),
\end{equation}
where $\mu\in[0,1]$ and the continuous function $g_\mu:\RR\ra [0,1]$ is defined as follows:
\begin{equation}\label{function gmu}
g_\mu(x)=\left\{\begin{array}{ll} g(x), & x<L(1+\mu), \\
\mu, & x\geq L(1+\mu).
\end{array}\right.
\end{equation}
Obviously, $\tH(0)=H(0)$ and $\tH(1)= H^s$. Also, if $p<L$, then $\tH_p(\mu)=H_p(0)$ regardless of the value of $\mu$, while for $p>2L$ $\tH_p(\mu)=H_p(\mu)$. By our basic assumption, the Hamiltonian $\tH(\mu)$ is gapped for all $\mu$, has a unique ground state,  and its correlation length is of order $\xi$. We can write:
\begin{equation}
Q^s=\sum_q h(q)\int_0^\mu d\mu \frac{d}{d\mu} \langle Q_q\rangle_\mu  +O(L^{-\infty}),
\end{equation}
where $\langle\ldots\rangle_\mu$ denotes the expectation value in the ground-state corresponding to $\tH(\mu).$ 

On the other hand, for the one-parameter family $\tH(\mu)$ we have an identity
\begin{equation}
d\langle Q_q\rangle_\mu=\sum_p \tT^{(1)}_{pq}.
\end{equation}
Here the 1-form $\tT^{(1)}_{pq}$ on $[0,1]$ is given by 
\begin{equation}
\tT^{(1)}_{pq}=\oint \frac{dz}{2\pi i} {\rm Tr}\left(\tG d\tH_p \tG Q_q-\tG d\tH_q \tG Q_p\right),
\end{equation}
and $\tG=(z-\tH)^{-1}.$ It is skew-symmetric under the interchange of $p,q$ and decays exponentially when $|p-q|$ is large. Using this, we can re-write the skyrmion charge as follows:
\begin{equation}\label{tildeQh}
Q^s=\frac12 \int_0^1 \sum_{pq} (h(q)-h(p)) \tT^{(1)}_{pq}+O(L^{-\infty}).
\end{equation}
Now note that since the function $h(x)=\theta(x)-\theta(x-3L)$ is constant on the scale $\xi$ everywhere except near $x=0$ and $x=3L$, only the neighborhoods of $p=q=0$ and $p=q=3L$ may  contribute appreciably to the double sum over $p,q$. One can make this explicit by writing
\begin{equation}
Q^s=Q^s_0-Q^s_{3L}+O(L^{-\infty}),
\end{equation}
where $Q^s_a$ is obtained from the r.h.s. of eq. (\ref{tildeQh}) by replacing $h(x)$ with $\theta(x-a)$. 

Now, since $\tH_p(\mu)$ is independent of $\mu$ for $p<L$, the 1-forms $\tT^{(1)}_{pq}$ are identically zero when both $p$ and $q$ are in the neighborhood of $x=0$. Therefore $Q^s_0$ is exponentially small for $L\gg\xi$. Further, when both $p$ and $q$ are in the neighborhood of $x=3L$, the 1-forms $\tT^{(1)}_{pq}$ are exponentially close to the 1-forms $T^{(1)}_{pq}$ for the Hamiltonian $H(\mu)$. This follows from the insensitivity of the correlators of the form
\begin{equation}\label{KubopairingAB}
\oint \frac{dz}{2\pi i}{\rm Tr} (G A G B)
\end{equation}
to the Hamiltonian far from the support of $A$ and $B$ \cite{Watanabe}. 
Thus
\begin{equation}
Q^s=-\frac12 \int \sum_{p,q}(\theta(q-3L)-\theta(p-3L))T^{(1)}_{pq}+O(L^{-\infty}).
\end{equation}
Comparing with eq. (\ref{Q1d}) and taking the limit $L\ra\infty$, we conclude that the skyrmion charge is minus the value of the  Thouless charge pump for the corresponding loop. 

Now we use the same approach to understand the physical meaning of the Thouless charge pump invariant for 2d systems. We start with a family of $U(1)$-invariant gapped 2d Hamiltonians parameterized by $S^1\times S^1$:
\begin{equation}
H(\lambda,\sigma)=\sum_p H_p(\lambda,\sigma).
\end{equation}
Here $\lambda$ and $\sigma$ are periodically identified with period $1$. We are going to associate to this two-parameter family two one-parameter families. The first one is simply
\begin{equation}
H^0(\lambda)=\sum_p H_p(\lambda,0).
\end{equation}
To define the second one, we need to choose a strip $\cS$ on $\RR^2$ of width $3L$ much larger than the correlation length. We choose coordinates on $\RR^2$ so that the strip is given by the inequalities $0\leq y\leq 3L$. Let us also denote by $\cS_0$ the strip of width $L$ which in this coordinate system is given by $L\leq y\leq 2L$. Obviously, $\cS_0\subset \cS.$ We define
\begin{equation}
H^s(\lambda)=\sum_p H^s_p(\lambda)=\sum_p H_p(\lambda,g(y(p))),
\end{equation}
where $g$ is defined in (\ref{function g}).
Thanks to the periodicity in $\sigma$, $H^s_p(\lambda)$ coincides with $H_p(\lambda,0)$ outside of the strip $\cS_0$. Also, both families are $U(1)$-invariant and periodic in $\lambda$ with period $1$. 
By our basic technical assumption, for sufficiently large $L$ the Hamiltonians $H^s(\lambda)$ are gapped for all $\lambda$. 

Now consider adiabatically varying $\lambda$ as a function of time. As one varies $\lambda$ from $0$ to $1$, the charge flows across the line $x=0$. The net charge transport across $x=0$ will be infinite both for $H^0(\lambda)$ and $H^s(\lambda)$. However, their difference is finite. This is because outside the strip $\cS_0$ the two Hamiltonians coincide, and thus outside the horizontal strip $\cS$ correlators of the form (\ref{KubopairingAB}) are the same up to terms which are exponentially suppressed far from $\cS$. The difference of net charges transported across $x=0$ is
\begin{equation}\label{doubledeltaQ}
\Delta Q^s-\Delta Q^0=\frac12\int_0^1\sum_{pq}(f(q)-f(p))\left(T^{s\,(1)}_{pq}(\lambda)-T^{(1)}_{pq}(\lambda,0)\right),
\end{equation}
where $f(p)=\theta(x(p))$, and $T^{s\,(1)}_{pq}(\lambda)$ and $T^{(1)}_{pq}(\lambda,0)$ are the 1-forms (\ref{T1}) on $[0,1]$ for Hamiltonian families $H^s(\lambda)$ and $H^0(\lambda)$, respectively. The above expression for $\Delta Q^s-\Delta Q^0$ can be interpreted in more physical terms by stacking the family $H^s(\lambda)$ with the time-reversal of the family $H^0(\lambda)$. This gives a family of 2d systems for which the Thouless charge pump across the line $x=0$ is finite, because the charge transport cancels out outside the strip $\cS$.

Next we introduce a cut-off function $h(p)=\theta(y(p))-\theta(3L-y(p))$ and write
\begin{equation}
\Delta Q^s-\Delta Q^0=\frac12\int_0^1\sum_{pq}(f(q)-f(p))h(q)\left(T^{s\,(1)}_{pq}(\lambda)-T^{(1)}_{pq}(\lambda,0)\right)+O(L^{-\infty}).
\end{equation}
Using the chain-cochain notation, this can also be written as
\begin{equation}\label{doubledeltaQh}
\Delta Q^s-\Delta Q^0=\int_0^1\langle T^{s\,(1)}_{pq}(\lambda),\delta f\cup h\rangle-\int_0^1 \langle T^{(1)}(\lambda,0),\delta f\cup h\rangle+O(L^{-\infty}).
\end{equation}
The advantage of introducing the cut-off function $h$ is that now both terms in eq. (\ref{doubledeltaQh}) are separately well-defined.

To compute the r.h.s. of (\ref{doubledeltaQh}) we  construct a two-parameter family of gapped Hamiltonians $\tH(\lambda,\mu)$ which interpolates between the family  $H^s(\lambda)$ and the family $H^0(\lambda)$. We define
\begin{equation}
\tH(\lambda,\mu)=\sum_p \tH_p(\lambda,\mu)=\sum_p H_p(\lambda,g_\mu(y(p))),
\end{equation}
where $g_\mu:\RR\ra[0,1]$ is defined by eq. (\ref{function gmu}). After the same kind of manipulations that lead from (\ref{tildeQ}) to (\ref{tildeQh}) we get
\begin{equation}
\Delta Q^s-\Delta Q^0 =-\int \langle \tT^{(2)},\delta f\cup\delta h\rangle+O(L^{-\infty}),
\end{equation}
where the integration is over the square $[0,1]^2$ in the $\lambda-\mu$ plane. The contraction of a 2-chain with a 2-cochain involves a triple sum over $p,q,r\in\Lambda$. Now we note that only the terms where all three points $p,q,r$ are close to the lines $y=3L$ or $y=0$ contribute appreciably to the sum.
The contribution of $y=0$ is of order $O(L^{-\infty})$, because $\tH_p(\lambda,\mu)$ does not depend on $\mu$ there and thus the 2-form $\tT^{(2)}_{pqr}$ is exponentially small. When evaluating the contribution of $y=3L$, one can replace $\tT^{(2)}$ with $T^{(2)}$ while making an error of order $O(L^{-\infty}).$ Thus we get
\begin{equation}
\Delta Q^s-\Delta Q^0 =\int \langle T^{(2)},\delta f\cup\delta h_{3L}\rangle+O(L^{-\infty}),
\end{equation}
where $h_{3L}(p)=\theta(y(p)-3L)$. 
Taking the limit $L\ra\infty$ we conclude that $\Delta Q^s-\Delta Q^0$ is the topological invariant of the family $H(t,\sigma)$.

\section{Higher Thouless charge pump for systems of free fermions in 2d}\label{appednix: free fermions}

In this section we compute the descendant of the Thouless charge pump for a family of systems of free fermions in two spatial dimensions (that is, for families of 2d insulators of class A). The two-form $T^{(2)}_{p_0p_1p_2}$ can be found from (\ref{tn expr}) to be 
\begin{align}\label{appendix:t2}
    T^{(2)}_{p_0p_1p_2} =\sum_{\sigma \in {\mathcal S}_{3}} (-1)^{{\rm sgn}\, \sigma}\oint \frac{dz}{2\pi i} \Tr \left( G dH_{p_{\sigma(0)}}G dH_{p_{\sigma(1)}} G Q_{p_{\sigma(2)}} \right).
\end{align}
We will compute this expression for the following many-body Hamiltonian density
\begin{equation}
H_p=\frac12\sum_{m\in\Lambda} \left(a^\dagger_ph(p,m)a_m+a^\dagger_m h(m,p) a_p\right),
\end{equation}
where $h(p,q)$ is an infinite Hermitian matrix $h(p,q)^*=h(q,p)$ whose rows and columns are labeled by $\Lambda$, and $a_p,a^\dagger_p$, $p\in\Lambda,$ are fermionic operators satisfying anticommutation relation 
\begin{align}
    \begin{split}
          \{ a^\dagger_p, a_q\}&= \delta_{p,q},\\
            \{ a_p, a_q\}&=      \{ a^\dagger_p, a^\dagger_q\}=0,
      \end{split}
\end{align}
with $\delta_{p,q}$ being the Kronecker delta. We define the conserved charge density to be
\begin{equation}
    Q_p = a^\dagger_p a_p.
\end{equation}

Expanding the many-body operators in eq.  (\ref{appendix:t2}), we find that it can be reduced to a single-particle correlation function 
\begin{align}
    T^{(2)}_{p_0p_1p_2} =\sum_{\sigma \in {\mathcal S}_{3}} (-1)^{{\rm sgn}\, \sigma}\oint \frac{dz}{2\pi i} \tr \left( g dh_{p_{\sigma(0)}}gdh_{p_{\sigma(1)}} g \delta_{p_{\sigma(2)}} \right),
\end{align}
where the contour of integration encloses filled states below the Fermi level and the trace is taken over the single-particle Hilbert space $\ell^2(\Lambda)$. Lowercase letters denote single-particle operators acting on the single-particle Hilbert space and functions on $\Lambda$ are thought of as multiplication operators on $\ell^2(\Lambda)$. In particular $h$ should be thought as an operator with matrix elements $h(p,q)$, its resolvent is $g=(z-h)^{-1}$, the Hamiltonian density is $h_p= \frac  1 2 (\delta_p h + h\delta_p)$, and the charge density operator is the Kronecker delta function $\delta_p$ which is equal to 1 on cite $p$ and 0 on all other cites. 

After contracting this expression with the 2-cochain $\delta f_1 \cup \delta f_2$, we find
\begin{align} \label{TP free fermion}
\begin{split}
   \langle T^{(2)},\delta f_1 \cup \delta f_2\rangle=  \frac 1 4\oint\frac{dz}{2\pi i} {\rm tr} \Big(2g^2[h,f_1]gdhg[h,f_2]gdh+g^2 [h,f_1]g[dh,f_2]gdh\\- g [dh,f_1]g[h,f_2]g^2 dh - (f_1 \leftrightarrow f_2)\Big ).
\end{split}
\end{align}
Both $f$ and $h$ are not trace class operators and it is not obvious that trace in the above expression exists. If we choose $f_1(p)= \theta(x^1(p))$ and $f_2(p)= \theta(x^2(p))$, the operators $[h,f_1]$ and $[h,f_2]$ are supported on vertical and horizontal lines with a finite intersection. Since the matrix elements of $g=(z-h)^{-1}$ decay faster than any power of the distance, the trace is convergent and well-defined. 

Up to this point we have not imposed any additional condition on the Hamiltonian besides being free and charge-conserving. If in addition we choose $\Lambda=\ZZ^2$ and impose a symmetry under lattice translations, than there is another natural topological invariant one can attach to a two-parameter family of such systems. Consider a translationally-invariant fermionic system depending on two parameters $\lambda_1,\lambda_2$ which are local coordinates on some closed surface $\Sigma$. We assume that the gap between valence bands and conduction bands does not close for any values of the parameters. Then Bloch wavefunctions of the valence bands form a vector bundle over the product of the Brillouin zone $S^1\times S^1$ and the parameter space $\Sigma$. One can define a non-Abelian Bloch-Berry connection with curvature $\mathcal F$ on this bundle. The integral of the degree 4 component of the Chern character of this connection over the product of Brillouin zone and the parameter space 
\begin{equation}
\int_{S^1\times S^1\times\Sigma} {\rm Ch}(\cf)=-\frac 1 {8\pi^2}\int_{S^1\times S^1\times\Sigma}  \Tr(\cf\wedge \cf)
\end{equation}
is a topological invariant of this family. 

This invariant can be expressed as non-linear response coefficient (see Sec. IIIA in \cite{QiHughesZhang}) 
\begin{align}
\begin{split}
    -\frac 1 {8\pi^2}\int_{S^1\times S^1\times\Sigma}  \Tr(\cf\wedge \cf) &= \frac{\pi^2}{15} \epsilon^{\mu\nu\rho\sigma\tau} \oint \frac{dz}{2\pi i} \int_{S_1\times S_1} \frac{d^2k}{(2\pi)^2}\int_{\Sigma_2} \frac{d^2\lambda}{(2\pi)^2}\\& {\rm tr'}\left[ \left(g \frac{\partial g^{-1}}{\partial q^\mu}\right)\left(g \frac{\partial g^{-1}}{\partial q^\nu}\right)\left(g \frac{\partial g^{-1}}{\partial q^\rho}\right)\left(g \frac{\partial g^{-1}}{\partial q^\sigma}\right)\left(g \frac{\partial g^{-1}}{\partial q^\tau}\right)\right],
\end{split}
\end{align}
where $q^\mu=(z,k_1,k_2 ,\lambda_1,\lambda_2)$, the first integral encloses the energies of the valence bands, the second integral is over the Brillouin zone, and the last integral is over the two-dimensional parameter space $\Sigma$. Here the trace $\tr'$ is taken over the space of valence-band Bloch wavefunctions with a fixed quasi-momentum. In order to relate this formula to (\ref{TP free fermion}) we should interprete the integral over the Brillouin zone as a part of the trace $\tr$ and replace $ \frac{\partial g^{-1}}{\partial k_i}$ with $i[h,f_i]$. We find
\begin{align}
 \begin{split}
-\frac 1 {8\pi^2}\int_{S^1\times S^1\times\Sigma}  \Tr(\cf\wedge \cf)=  -\frac{1}6 \oint \frac{dz}{2\pi i} \int_{\Sigma_2} {\rm tr}\Big[ g^2[h,f_1]g[h,f_2]gdhgdh\\-g^2[h,f_1]gdhg[h,f_2]gdh+g^2[h,f_1]gdhgdhg[h,f_2]+g^2dhg[h,f_1]g[h,f_2]gdh\\-g^2dhg[h,f_1]gdhg[h,f_2]+g^2dhgdhg[h,f_1]g[h,f_2]   - (f_1 \leftrightarrow f_2)\big],
  \end{split}
\end{align}
where  $dh =\sum_\ell\dfrac{\partial h}{\partial \lambda^\ell}d\lambda^\ell $. The integrand of this expression differs from (\ref{TP free fermion}) by a total derivative
\begin{align}
    \frac 1 {12} d\Big[\oint\frac{dz}{2\pi i} {\rm tr} \Big(g^2 [h,f_1]g[h,f_2]gdh- g [h,f_1]g[h,f_2]g^2dh - (f_1 \leftrightarrow f_2)\Big )\Big].
\end{align}
Therefore for the free fermions the 2d Thouless charge pump integrated over $\Sigma$ is equal to $\int_{S^1\times S^1\times\Sigma} {\rm Ch}(\cf)$. Since $\Sigma$ is arbitrary, this proves that the cohomology class of $Q^{(2)}(f_1,f_2)$ is  the cohomology class of the degree-4 compionent of the Chern character integrated over the Brillouin zone. 

One can construct an example with a non-trivial $Q^{(2)}$ by taking the 4d Chern insulator (see sec. IIIB of \cite{QiHughesZhang}) and declaring two components of the quasi-momentum to be parameters. The parameter space is a torus $T^2$ in this case. In this way one gets the following Hamiltonian: 
\begin{align}
    H=\sum_{k_x,k_y}\psi^\dagger_{\vec k} d_a(\vec k,{\vec \lambda}) \Gamma^a \psi_{\vec k},
\end{align}
where $\Gamma^a$ are five-dimensional Dirac matrices satisfying the Clifford algebra and 
\begin{equation}
    d_a(\vec k,{\vec \lambda})  = \left[(m+c+\cos k_x+\cos k_y+\cos \lambda_1+\cos \lambda_2), \sin k_x, \sin k_y, \sin \lambda_1,\sin \lambda_2\right].
\end{equation}
It was shown in \cite{QiHughesZhang} that for a particular choice of $m$ and $c$ this model has a  nontrivial integral of the Chern class ${\rm Ch}(\cf)$ over $T^4=T^2\times T^2$. 

\section{Discussion}

We have shown how to attach topological invariants to $D$-parameter families of $U(1)$-invariant gapped Hamiltonians in $D$ dimensions. In agreement with field theory, we found that they can be encoded in a closed $D$-form on the parameter space. The form itself depends on some additional choices, but its cohomology class does not. Thus integrals of this $D$-form over $D$-cycles in parameter space are independent of any choices. 

All these topological invariants arise from the ground-state charge via the "descent" equations proposed by Kitaev \cite{Kitaev_talk,Kitaev_unpublished}. This is analogous to how the Berry curvature of 0d systems gives rise to Wess-Zumino-Witten-type invariants of families of $D$-dimensional gapped systems without symmetries \cite{HigherBerry}. In general, one expects that every topological invariant of a gapped Hamiltonian in dimension $D$ gives rise to a topological invariant of a $k$-dimensional family of gapped Hamiltonians in dimension $D+k$. In particular, Hall conductance for 2d gapped systems with a $U(1)$ symmetry gives rise to a topological invariant of one-parameter families of 3d gapped systems with a $U(1)$ symmetry. We plan to discuss this invariant in a future publication.  

Topological invariants of families of gapped systems of free fermions were previously discussed by Teo and Kane \cite{TeoKane}. They interpreted the parameter space as the complement of a gapless defect in an otherwise gapped system. Thus the parameter space was homotopically equivalent to a sphere of dimension $D-k-1$, where $k$ is the dimension of the defect. Note that the dimension of the parameter space is strictly less than $D$, so this situation is distinct from the one considered here. But Ref. \cite{TeoKane} also considered a situation where the Hamiltonian depends on an additional periodic parameter $t$, so that the total parameter space is $S^{D-k-1}\times S^1$. For $k=0$ (point-like defects) this has dimension $D$ and thus can be compared with our construction. Indeed, for systems of class A Ref. \cite{TeoKane} assigns to such a family an integer topological invariant and interprets it as the net charge pumped towards the defect as the system undergoes an adiabatic cycle in the variable $t$. From the effective field theory perspective, considering a point-like defect in $\RR^D$ is the same as compactifying the system on $S^{D-1}$. Then the topological invariant discussed in Ref. \cite{TeoKane} is the Thouless charge pump of the resulting 1d system. One can regard the results obtained in this paper as a generalization of Ref. \cite{TeoKane} in two distinct directions: to the case of interacting $U(1)$-invariant systems and to the case of $D$-dimensional parameter spaces of arbitrary topology. The interpretation in terms of point-like defects is lost when one consider parameter spaces which are not of the form $S^{D-1}\times S^1$. It is desirable to find a physical  interpretation of the higher Thouless charge pump invariant  for arbitrary parameter spsces. Field theory suggests that it can be interpreted as the ground-state charge of the system compactified on a topologically non-trivial space and deformed by  spatially-varying parameters. Unfortunately, it is difficult to make sense of this in the world of lattice models, with the exception of the case when the spatial manifold is a torus.

\appendix
\section{Quantization of the Thouless charge pump and its descendants} \label{appendix:quantization}

Consider a family of gapped $U(1)$-invariant systems in $D$ dimensions parameterized by a $D$-dimensional sphere. Suppose also that all systems in the family are in a Short-Range Entangled (SRE) phase. In this appendix we argue that for such a family the descendant Thouless charge pump invariant is an integer. 

We start with the ordinary Thouless charge pump $Q^{(1)}(f)$ for gapped 1d systems. As far as we know, its integrality for interacting systems  was shown only recently \cite{Bachmannetal}. We will provide an alternative argument for it. Our argument is a slight modification of a similar one in Appendix A of Ref. \cite{HigherBerry} which shows integrality of the higher Berry curvature invariant for 1d lattice systems. Let us first sketch the rough idea. Due to a  finite correlation length of gapped systems the 1-form $Q^{(1)}(f)$ depends only on the static linear response in the neighborhood of the region where the function $f$ is non-constant.  For the  specific choice $f(p)= \theta(p-a)$ one can see from (\ref{Q1d}) that only points which are a few correlation lengths away from $a$ contribute significantly to $Q^{(1)}(f)$. Naively, one could expect that if we terminate the system away from point $a$  by replacing the rest of the Hamiltonian with a trivial one it would not change the value of $Q^{(1)}(f)$. The flaw  of this procedure is that the modified Hamiltonian is not guaranteed to be gapped for all values of the parameters. If for some values of parameters the gap closes, the correlation length might diverge and this might lead to additional contributions to  $Q^{(1)}(f)$ which depend on the behavior of the system far from the point $a$. Moreover, if a termination without gapless edge modes were possible everywhere in the parameter space, it would prove that the 1-form $Q^{(1)}(f)$ is exact via (\ref{q1 compactly supported}), and its integral over any cycle would be zero. Thus a nonzero value for periods of $Q^{(1)}(f)$ is an obstruction for finding a gapped termination which varies continuously with parameters.

Despite this, we will argue below  that if we focus on a particular loop in the parameter space and cover it with two semi-circles, it is possible to find a family of gapped terminations for each of the two semi-circles separately. A termination is constructed via a choice of a path in the parameter space and an adiabatic deformation (i.e. a slow spatial variation) of the original Hamiltonian on a half-line. Since the loop is assumed to be non-contractible, one cannot choose the path in the parameter space so that it depends continuously on the starting point and is defined everywhere on the loop. But it can be chosen continuously for any contractible part of the loop, such as the two semi-circles. We define via this procedure two families of gapped boundary conditions: one on the far right of the lattice (using the first semi-circle) and another one on the far left of the lattice (using the second semi-circle).  The modifications made to the family would not affect the value of $Q^{(1)}(f)$ and one can show that result of integration reduces to the charge of the finite system on the equator of the cycle (i.e. two points separating the semi-circles) which is integrally quantized.

After this rough exposition, let us describe the argument in more detail. Consider a loop $\phi:S^1 \rightarrow \cM$ in the parameter space $\cM$. Let $f(p)=\theta(p)$. We will show that 
\begin{equation}\label{appednix:q1 over cycle}
    \int_{S^1} \phi^*\big(Q^{(1)}(f)\big)  \in \mathbb Z.
\end{equation}
There is no topological order for 1d systems and thus all gapped systems of the family are in the same SRE phase. For gapped 1d systems with $U(1)$ symmetry, there is only one SRE phase: the trivial one. Therefore all systems in the family are in the trivial phase.

Let $\cm_0 \in \cM$ be some specific system in the trivial phase. That is, $\cm_0$ is some particular  gapped Hamiltonian $H=\sum_p H_p$ where each $H_p$ acts only on site $p$, is $U(1)$-invariant, and has a unique ground state. Each point in the image of $\phi$ can be connected to $\cm_0$ by a continuous path in $\cM$. But since the loop $\phi$ is assumed to be non-contractible, one cannot choose these paths for all points on $S^1$ in a continuous fasion. Let us delete the north (resp. south) pole of $S^1$ and call the resulting subset $S^1_S$ (resp. $S^1_N$). Since they are contractible, their images can be continuously deformed to $\cm_0$, and we denote the corresponding homotopies by $\mathcal P_S$ and $\mathcal P_N$. These are continuous maps from $[0,1]\times S^1_S$ to $\cM$ and from $[0,1]\times S^1_N$ to $\cM$, respectively. They satisfy $\mathcal P_S(0,\lambda)=\phi(\lambda) $ and  $\mathcal P_N(0,\lambda)=\phi(\lambda)$ and $\mathcal P_S(1,\lambda)=\cm_0 $ and  $\mathcal P_N(1,\lambda)=\cm_0$. The parameter $\lambda$ takes values in $S^1$.

Denote the Hamiltonian corresponding to a point $\cm\in\cM$ by $H(\cm)=\sum_p H_p(\cm)$. The Hamiltonian corresponding to $\lambda\in S^1$  is $H[\lambda] = \sum_p H_p(\phi(\lambda))$. For a point $\lambda\in S_N^1$ of the circle we define a new Hamiltonian $H_p^+[\lambda]= H_p(\cm(\lambda,p))$ where we let the parameters of the Hamiltonian to depend adiabatically on the site $p$ as $\cm(\lambda,p)=\mathcal P_N(t_N(p),\lambda)$. The function $t_N:\RR\ra [0,1]$ is equal to 1 for $p\in [2L,+\infty)$, smoothly interpolates from 1 to 0 in the region $p\in [L,2L]$, and is 0 for $p\in (-\infty,L]$. Similarly, we define  another local Hamiltonian $H^-[\lambda]$ for all points $\lambda\in S^1_S$ as $H^-[\lambda]=\sum_{p\in\Lambda} H_p(P_S(t_S(p),\lambda))$ where the function $t_S:\RR\ra [0,1]$ is equal to 1 for $p\in (-\infty,-2L],$ smoothly interpolates between 1 to 0 for $p\in [-2L,-L],$ and is 0 for $p\in [-L,+\infty)$. The last Hamiltonian we define is $H^{+-}_p[\lambda]$ for all point $\lambda$ in the intersection $\in S^1_N\bigcap S^1_S$ which coincides with $H_p[\lambda]$ for $p\in [-L,L]$, coincides with $H^+_p[\lambda]$ for $p\in [L,+\infty)$, and coincides with $H^-_p[\lambda]$ for $p\in (-\infty,-L].$ 

In order for our argument to work, we need to assume that all these Hamiltonians are gapped for sufficiently large $L$. Since the correlation lengths of the Hamiltonians $H[\lambda]$ are bounded from above and $L$ can be taken much larger than this bound, this assumption seems very plausible.  Let $Q^{(1)}_+(f)$, $Q^{(1)}_-(f)$ and $Q^{(1)}_{+-}(f)$ be the 1-forms corresponding to the families $H^+$, $H^-$ and $H^{+-}$, respectively. They are defined on $S^1_N$, $S^1_S$ and $S^1_N\bigcap S^1_S$, respectively. We separate the integral over $S^1$ into a sum of integrals over the north and south semi-circles $B_+$ and $B_-$:
\begin{equation*}
    \int_{S^1} \phi^*(Q^{(1)}(f) )= \int_{B_+} \phi^*(Q^{(1)}(f))+ \int_{B_-} \phi^*(Q^{(1)}(f)) =\int_{B_+} Q^{(1)}_+(f)  +\int_{B_-} Q^{(1)}_-(f)+ O(L^{-\infty}),
\end{equation*}
where in the last step we replaced $\phi^*(Q^{(1)}(f))$ with $Q^{(1)}_\pm(f)$. By our assumption, all three  Hamiltonians $H[\lambda],H^+[\lambda],$ and $H^-[\lambda]$ are gapped, and the 1-form $\phi^*(Q^{(1)}(f))$ is sensitive only to the behavior of the system in a neighborhood of the point $p=0$ where  the function $f(p)=\theta(p)$ is non-constant. These three families differ from each other only outside of the interval $p\in[-L,L]$, and for large $L$ the difference between these 1-form is of order $L^{-\infty}$.

Next we define $f_+(p)=\theta(p-3L)$ and $f_-(p)= \theta(p+3L)$ and write
\begin{equation}
 \int_{B_+} Q^{(1)}_+(f)  +\int_{B_-} Q^{(1)}_-(f)=\int_{B_+} Q^{(1)}_+(f_+)  +\int_{B_-} Q^{(1)}_-(f_-)+\int_{B_+} Q^{(1)}_+(f-f_+)  +\int_{B_-} Q^{(1)}_-(f-f_-).
\end{equation}
Near the point $p=3L$ the Hamiltonian $H^+[\lambda]$ coincides with the constant Hamiltonian $H(\cm_0)$ and therefore the form $ Q^{(1)}_+(f_+)$ as well as its integral over $B_+$ is of order $L^{-\infty}$. Similarly, the integral $\int_{B_-} Q^{(1)}_-(f_-)=O(L^{-\infty})$. The remaining two terms contain compactly supported functions $f_\pm -f $. For any function $g$ with compact support one can use the  equation (\ref{q1 compactly supported}) which gives $Q^{(1)}_\pm(g) =\langle T^{(1)}_\pm,\delta g\rangle = d \langle T^{(0)}_\pm, g\rangle$. Here $T^{(0)}_\pm$ is the 0-chain with components $\langle Q_p\rangle $. It depends on the Hamiltonian through the ground state. Therefore we find
\begin{equation}
\int_{B_+} Q^{(1)}_+(f-f_+)  +\int_{B_-} Q^{(1)}_-(f-f_-) =  \langle T^{(0)}_+(\lambda_e)-T^{(0)}_+(\lambda_w),f-f_+\rangle  -\langle T^{(0)}_-(\lambda_e)-T^{(0)}_-(\lambda_w),f-f_-\rangle,
\end{equation}
where $\lambda_e,\lambda_w$ are the two points constituting the common boundary of $B_+$ and $B_-$, and $T^{(0)}_\pm(\lambda)$ are computed for the corresponding Hamiltonians at these points. The signs in this formula are defined by the orientations of the boundaries. Now we can  replace $T^{(0)}_+$ and $T^{(0)}_-$  with $T^{(0)}_{+-}$ in both sums. Such a replacement introduces an error of order $L^{-\infty}$, because the summands are only sensitive to the the Hamiltonian of the systems in the region where $H^+_p[\lambda]=H^{+-}_p[\lambda]$ and $H^-_p[\lambda]=H^{+-}_p[\lambda]$.
This gives
  
\begin{align}\label{finalT}\begin{split}
 \langle T^{(0)}_+(\lambda_e)&-T^{(0)}_+(\lambda_w),f-f_+\rangle  -\langle T^{(0)}_-(\lambda_e)-T^{(0)}_-(\lambda_w),f-f_-\rangle\\&=  \langle T^{(0)}_{+-}(\lambda_e)-T^{(0)}_{+-}(\lambda_w),f-f_+\rangle  -\langle T^{(0)}_{+-}(\lambda_e)-T^{(0)}_{+-}(\lambda_w),f-f_-\rangle+O(L^{-\infty})  \\&= - \langle T ^{(0)}_{+-}(\lambda_e)-T^{(0)}_{+-}(\lambda_w),f_+-f_-\rangle  +O(L^{-\infty}). 
\end{split}
\end{align}
Recall that by construction when $p\notin [-2L,2L]$ the Hamiltonian $H^{+-}_p[\lambda]=H_p(\cm_0)$ is a collection of decoupled $U(1)$-invariant Hamiltonians. In this region $\langle Q_p\rangle $ is integral and independent of $\lambda$, so the contributions of $\lambda_e$ and $\lambda_w$ to eq. (\ref{finalT}) cancel for each $p$ separately. Thus the sum over $p$ implicit in eq. (\ref{finalT}) can be restricted to $p\in[-2L,2L]$. Since the region $p\in [-2L,2L]$ is decoupled from the rest of the system, we are dealing with a finite size system of length $4L$ (or more precisely, two such systems corresponding to points $\lambda_e$ and $\lambda_w$). Further, the function  $f_+(p)-f_-(p)= \theta(p-3L)-\theta(p+3L)$ is equal to $-1$ in this region, so the evaluation of the chain on a cochain in (\ref{finalT}) is the difference of ground-state charges of two finite systems corresponding to parameters $\lambda_w$ and $\lambda_e$. This difference is obviously integral.
Thus we showed that 
\begin{equation}
    \int_{S^1} \phi^*(Q^{(1)}(f) ) =n +O(L^{-\infty}), \quad n\in\mathbb Z.
\end{equation}
Taking the limit $L\rightarrow \infty$ we conclude that the Thouless charge pump invariant  for a gapped 1d system with a unique ground state is integral. 

To extend this argument to higher dimensions, we use induction in $D$. We need to assume that all  systems in the family are in a Short-Range Entangled phase. This is to be expected, since in the presence of topological order one expects topological invariants to become fractional. Other than that, the inductive step proceeds exactly as for $D=1$. First we reduce to the case of systems in a trivial phase by stacking the family under consideration with a fixed Short-Range Entangled system. Then we cover the parameter space $S^D$ with two contractible charts, $S^D_S$ and $S^D_N$, and on each of these charts construct a $D$-parameter family of Hamiltonians with adiabatically varying parameters which reduces to the trivial Hamiltonian for $x^D(p)\ll 0$ and $x^D(p)\gg 0$, respectively. We also construct a third family $H_{+-}$ which is defined on $S^D_N\bigcap S^D_S$ and interpolates to the trivial Hamiltonian both for $x^D(p)\gg 0$ and $x^D(p)\ll 0$. The same manipulations as for $D=1$ show that the invariant $Q^{(D)}(f_1,\ldots,f_D)$ is equal, up to terms of order $L^{-\infty}$, to the invariant $Q^{(D-1)}(f_1,\ldots,f_{D-1})$ of a certain family of gapped $(D-1)$-dimensional systems occupying the region $x^D(p)\in [-2L,2L]$ in $\RR^D$ and parameterized by the equatorial $S^{D-1}$. This concludes the inductive step.

\bibliographystyle{apsrev4-1}
\bibliography{bib}

\end{document}